\documentclass[aps,pre,superscriptaddress,notitlepage,10pt]{revtex4-1}

\usepackage{amsmath,amssymb,amsfonts}
\usepackage{graphicx}
\usepackage[dvipsnames]{xcolor}
\usepackage{comment}
\usepackage[overload]{empheq}
\usepackage{upgreek}
\usepackage{lmodern,pdftexcmds}
\usepackage[sort&compress]{natbib}
\usepackage{hyperref}
\usepackage{soul}
\usepackage[normalem]{ulem}
\usepackage{enumerate}
\usepackage{float}

\hypersetup{
    bookmarksopen=true,
    unicode=false,          
    pdftoolbar=true,        
    pdfmenubar=true,        
    pdffitwindow=false,     
    pdfstartview={Fit},    
    pdftitle={},    
    pdfauthor={},     
    colorlinks=true,       
    linkcolor=black,          
    citecolor=black,        
    urlcolor=black,           
    pdfborder={0 0 0}
}

\newcommand{\beq}{\begin{equation}}
\newcommand{\eeq}{\end{equation}}
\newcommand{\beqa}{\begin{eqnarray}}
\newcommand{\eeqa}{\end{eqnarray}}

\renewcommand{\bm}[1]{\boldsymbol{\mathbf{#1}}}

\providecommand*{\dd}[3][]{\frac{\mathrm{d}^{#1}#2}{\mathrm{d} #3^{#1}}}

\newcommand{\qmbox}[1]{\quad \mbox{#1} \quad}

\providecommand*{\rmd}{\mathrm{d}}

\providecommand*{\bm}{\mathbf}

\newcommand{\K}{\bm{{K}}}
\newcommand{\N}{\bm{{N}}}

\newcommand{\Jpf}{\bm{\mathcal{J}}}
\newcommand{\Kpf}{\bm{\mathcal{K}}}

\begin{document}

\preprint{}

\title{Motion of hydrodynamically interacting active particles}



\author{Bhargav Rallabandi}
\email{bhargav@engr.ucr.edu}
\affiliation{Department of Mechanical Engineering, University of California, Riverside, California 92521, USA }

\author{Fan Yang}
\affiliation{Department of Mechanical and Aerospace Engineering, Princeton University, Princeton, New Jersey 08544, USA}

\author{Howard A. Stone}
\email{hastone@princeton.edu \\}
\affiliation{Department of Mechanical and Aerospace Engineering, Princeton University, Princeton, New Jersey 08544, USA}

\date{\today}

\begin{abstract}
We develop a general hydrodynamic theory describing a system of interacting actively propelling particles of arbitrary shape suspended in a viscous fluid. We model the active part of the particle motion using a slip velocity prescribed on the otherwise rigid particle surfaces. We introduce the general framework for particle rotations and translations by applying the Lorentz reciprocal theorem for a collection of mobile particles with arbitrary surface slip. We then develop an approximate theory applicable to widely separated spheres, including hydrodynamic interactions up to the level of force quadrupoles. We apply our theory to a general example involving a prescribed slip velocity, and a specific case concerning the autonomous motion of chemically active particles moving by diffusiophoresis due to self-generated chemical gradients.
\end{abstract}


\maketitle 
\section{Introduction}
The self-propelled motion of particles suspended in viscous fluid has widespread applications in biological and synthetic systems. For microorganisms, achieving locomotion at small Reynolds numbers is inherently challenging due to the time-reversal symmetry of negligible-inertia hydrodynamics. Swimming microorganisms usually work around this limitation through non-reciprocal undulations of structures, with typical strategies including the propagation of ciliary or flagellar waves \citep{riedel2005self,guirao2007spontaneous,goldstein2015green,gilpin2017vortex}.

Achieving this type of synchronous mechanical actuation at small scales is challenging in synthetic systems. Researchers have instead sought to use chemical gradients to drive the motion of suspended particles with engineered surface properties. A particle whose surface interacts with the solute through a short-range potential will translate when exposed to a gradient of a chemical potential. This type of motion, known as diffusiophoresis \citep{derjaguin47}, can be generated by several physical mechanisms, including van der Waals forces, steric interactions or due to the electrostatic potential between a charged particle and ions in solution \citep{anderson82,anderson,JFM1984,sharifi13,velegol16}. Chemical concentration gradients necessary to drive motion in such synthetic systems may be applied either externally \citep{shin16,shi2016diffusiophoretic}, or may arise through the activity of the colloids themselves. Autonomous colloidal particles of the latter kind are typically designed with materials that catalyze chemical reactions, generating chemical species (and gradients thereof) that drive particle motion through phoretic mechanisms \citep{moran17}. Such synthetic autonomous motors have been realized experimentally with catalytic Janus colloids (usually containing silver, gold or platinum) \citep{paxton06,paxton05,sen09,moran11}, and have inspired modeling efforts \citep{gol05_catalyticmotor,udi11,wall,janus,tua18_artificial}. Directed motion of chemically active colloids may also occur as a result of geometric or dynamic symmetry breaking and has been studied both with theory \citep{twosphere,soto2014self,lauga} and experiments \citep{wykes2016dynamic}.

It is often effective to model the surface undulations of swimming microorganisms by prescribing a surface slip velocity distribution as an effective boundary condition on the otherwise rigid particle surface (the squirmer model) \citep{lighthill1952squirming,blake1971spherical,sto96_PRL_squirmer,ishikawa2006hydrodynamic}. A similar slip velocity also arises naturally as an effective condition for phoretic motion with thin interaction layers \citep{anderson}. In more coarse-grained approaches, the active particle is modeled as a finite superposition of force multipoles, which typically reproduces far-field characteristics of the fluid flow \citep{dre10_measurement}. This latter approach has nonetheless been effective at describing hydrodynamic interactions between swimmers and with surfaces in many systems \citep{lauga2006swimming,spa12_boundary,poo07_swimmer_interactions,ber08_attraction_surfaces,sin18_generalizedlaws} and has lent itself to continuum descriptions of active suspensions \citep{sai08_suspensions}. By representing body deformations using collections of rigid spherical particles, self-propulsion has also been modeled using the Stokesian dynamics framework  \citep{swa11_stokesian_swim}. Recently, Varma et al. \citep{var18_clustering_autophoresis} used a multipole expansion, complemented by boundary integral simulations, to show that identical phoretic spheres releasing chemical into the fluid can self-aggregate into stable clusters that can itself translate and rotate due to hydrodynamic interactions.  With the growing interest in the collective behavior of living active systems and the possibility of tailoring synthetic ones for applications such as drug delivery, it is useful to have a general framework to describe the hydrodynamics of interacting active particles.

The primary focus of this paper is to develop such a general hydrodynamic theory describing the motion of many interacting active particles with arbitrary surface velocity distributions. The paper is organized as follows. In section II, 
we first use the Lorentz reciprocal theorem to derive an exact formulation of the swimming speed, including all hydrodynamic interactions (within the Stokes flow approximation) for any $n$-particle system in terms of their geometry and surface slip velocities. Then, in section III, we apply this exact framework along with results from the hydrodynamics literature to develop approximations for widely separated spherical swimmers, accounting for hydrodynamic interactions up to and including force- and torque-quadrupoles. Section IV presents example calculations of the motion of a collection of particles with prescribed surface velocity, reminiscent of model swimming organisms, focusing first on the common situation of a multipolar surface slip. As a special case, we then analyze the motion of autophoretic chemically active particles that propel through geometric asymmetry, and recover the results of Varma et. al \citep{var18_clustering_autophoresis}. We demonstrate some effects of polydispersity in particle mobilities on the behavior of the system. We discuss the scope and some generalizations of the formalism and present conclusions in section \ref{SecConclusions}.

\section{General hydrodynamic theory} \label{SecLRT}
\begin{figure}
  \includegraphics[scale=0.9]{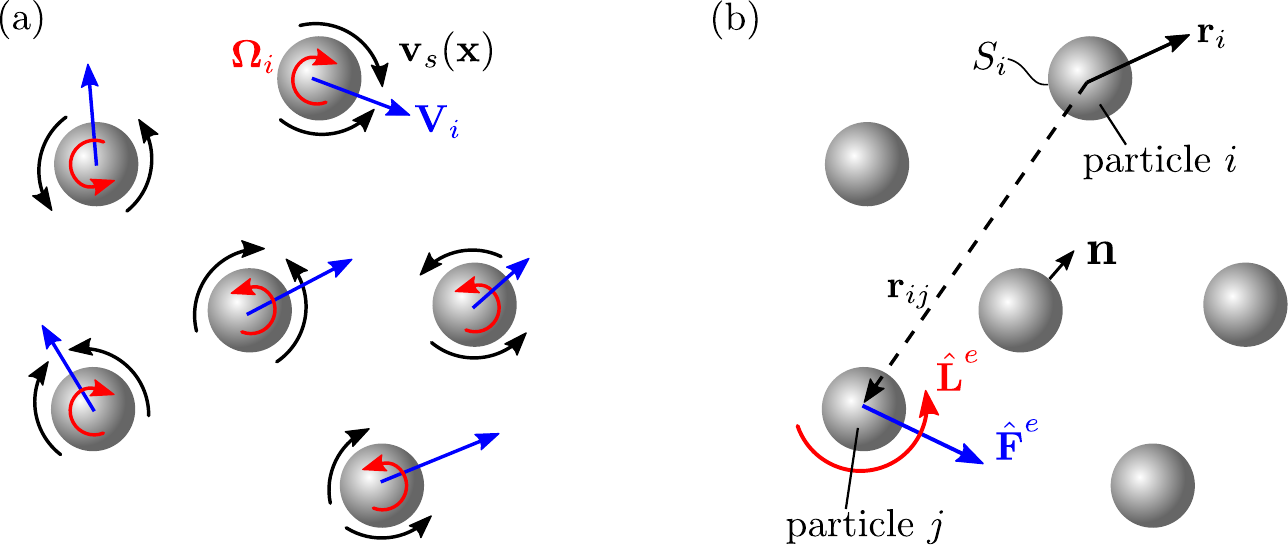}
\caption{Sketch of a system of $n$ active particles showing the setup of (a) the main problem and (b) the associated model problems. The main problem consists of $n$ force- and torque-free particles with prescribed surface velocities $\bm{v}_s(\bm{x})$, that translate with velocities $\bm{V}_i$ and rotate with angular velocities $\bm{\Omega}_i$. The quantities $\bm{V}_i$ and $\bm{\Omega}_i$ are calculated with the help of model problems where an external force $\hat{\bm{F}}^{e}$ or a torque $\hat{\bm{L}}^{e}$ acts on a single no-slip particle $j$, where all the other particles are force- and torque-free.}
\label{FigSketch}
\end{figure}

We consider the low-Reynolds-number motion of $n$ self-propelling particles of arbitrary shape suspended in Newtonian fluid of viscosity $\mu$. Each particle $i \in \{1,2,\dots,n\}$ translates with velocity $\bm{V}_i$ and rotates with angular velocity $\bm{\Omega}_i$ due to a combination of autonomously generated motions and, in general, externally applied forces and torques. Independent of the physical mechanism responsible for autonomous motion, we will adopt the common modeling framework wherein each particle $i$ is assumed rigid, but possesses a velocity distribution $\bm{v}_{s}(\bm{x} \in S_i)$ on its surface $S_i$ (Fig. \ref{FigSketch}a). Note that $\bm{v}_s(\bm{x})$ 
can implicitly vary in time and may depend the positions and velocities of all the other particles. Denoting the center of mass of particle $i$ by $\bm{x}_i$ and defining $\bm{r}_i = \bm{x} - \bm{x}_i$, the fluid velocity on its surface is therefore 
\begin{equation} \label{V_surface}
  \bm{v}(\bm{x}) = \bm{V}_i +  \bm{\Omega}_i \wedge \bm{r}_i + \bm{v}_s (\bm{x}) \qmbox{for} \bm{x} \in S_i\,.
\end{equation}
External forces $\bm{F}^e_i$ and torques $\bm{L}^{e}_i$ may additionally act on the particles. We note that active particles in many practical situations, particularly at the microscale, are nearly free of external forces or torques, although is not necessarily the case in all systems \citep{dre09_dancing,dre10_measurement}.

The surface velocity distributions together with the action of the applied forces and torques cause the particles to move and thereby disturb the surrounding fluid. We denote the fluid velocity and stress fields by $\bm{v}(\bm{x})$ and $\bm{\sigma}(\bm{x}) = - p \bm{I} + \mu (\nabla \bm{v} + ( \nabla \bm{v})^T)$, respectively, where $p(\bm{x})$ is the pressure. We note that all quantities may be implicitly time-dependent due to the changing configurations of the particles as they translate and rotate. Assuming negligible fluid inertia, the flow satisfies the Stokes equations 
\begin{align}
\nabla \cdot \bm{v} = 0 \qmbox{and} \nabla \cdot \bm{\sigma} = \bm{0}\,.
\end{align}
For particles with negligible inertia, the external forces and torques are balanced by their hydrodynamic counterparts, i.e. 
\begin{align} \label{FAndL}
\bm{F}^e_i = -\int_{S_i} \bm{n} \cdot \bm{\sigma} \,\rmd S \qmbox{and} \bm{L}^{e}_i = -\int_{S_i} \bm{r}_i \wedge (\bm{n} \cdot \bm{\sigma})\, \rmd S,
\end{align}
where $\bm{n}(\bm{x} \in S_i)$ is the unit normal to the particle surface $S_i$, directed towards the fluid (Fig. \ref{FigSketch}).

First, we use the Lorentz reciprocal theorem for Stokes flows to develop a general framework for arbitrary, hydrodynamically coupled, $n-$particle systems with a surface velocity given by \eqref{V_surface}. It is useful to introduce a model Stokes flow problem with velocity and stress fields $\hat{\bm{v}}(\bm{x})$ and $\hat{\bm{\sigma}}(\bm{x})$, respectively, which we define precisely in the following. The reciprocal theorem relates the main flow $(\bm{v}, \bm{\sigma})$ to the model flow ($\hat{\bm{v}}, \hat{\bm{\sigma}}$) by
\begin{align} \label{RT1}
  \sum_i  \int_{S_i} \bm{n} \cdot \hat{\bm{\sigma}} \cdot  \bm{v}\, \rmd S = \sum_i \int_{S_i} \bm{n} \cdot \bm{\sigma} \cdot  \hat{\bm{v}}\, \rmd S,
\end{align}
where the summations account for all the particle surfaces. The above expression is valid both when the flows in both problems decay far away from the particles, and when the particles are bound externally by rigid no-slip walls or by negligibly-deforming stress-free interfaces.  It is convenient to choose a model flow problem corresponding to the motion of rigid, no-slip particles with linear and angular velocities $\hat{\bm{V}}_i$ and  $\hat{\bm{\Omega}}_i$, respectively, under the action of externally applied forces $\hat{\bm{F}}^e_i$ and torques $\hat{\bm{L}}^e_i$ (Fig. \ref{FigSketch}b). Using the definitions \eqref{V_surface} and \eqref{FAndL} yields
\begin{align} \label{RT2}
  \sum_i  \hat{\bm{F}}^e_i \cdot \bm{V}_i + \sum_i \hat{\bm{L}}^e_i \cdot \bm{\Omega}_i =   \sum_i \bm{F}^e_i \cdot \hat{\bm{V}}_i + \sum_i \bm{L}^e_i \cdot \hat{\bm{\Omega}}_i + \sum_i  \int_{S_i} \bm{n} \cdot \hat{\bm{\sigma}} \cdot  \bm{v}_s \rmd S\,.
\end{align}
For a particular choice of model flow problem, \eqref{RT2} is a single scalar equation for $2n$ unknown vectors (and pseudovectors) $\bm{V}_i$ and $\bm{\Omega}_{i}$. The key to handling this seemingly underdetermined system is to recognize that \eqref{RT2} applies to \emph{any} choice of model problem corresponding to rigid translations and rotations of no-slip particles. 

While nearly any set of $2n$ model problems satisfying the properties discussed above can be used to compute $\bm{V}_i$ and $\bm{\Omega}_i$, a judicious choice results in considerable simplification. First, we introduce a set of $n$ model problems in which an external force $\hat{\bm{F}}^e$ acts on a single particle $j \in \{1,2,3,\ldots,n\}$, the remaining particles are force-free, and all $n$ particles (including particle $j$) are torque-free.  Denoting the flow field in this model problem by ($\hat{\bm{v}}^{F}_j(\bm{x})$, $\hat{\bm{\sigma}}^{F}_j(\bm{x})$), \eqref{RT2} reduces to 
\begin{align} \label{RT3}
  \hat{\bm{F}}^e \cdot \bm{V}_j  = \sum_i \bm{F}^e_i \cdot \hat{\bm{V}}_i + \sum_i \bm{L}^e_i \cdot \hat{\bm{\Omega}}_i  + \sum_i  \int_{S_i} \bm{n} \cdot \hat{\bm{\sigma}}^F_j \cdot  \bm{v}_s \rmd S ,
\end{align}
Recognizing that the force on particle $j$ is the only external input to the model problem, we invoke linearity of the Stokes equations to write $\hat{\bm{\sigma}}^F_j = \K_j (\bm{x})\cdot \hat{\bm{F}}^e$, where $\K_j(\bm{x})$ is the rank-3 tensor field that propagates the stress produced by an external force acting on particle $j$, and depends on the positions and orientations of all $n$ particles. Similarly, we write the velocity of any particle $i$ in the model problem as $\hat{\bm{V}}_i = \bm{M}^{VF}_{ij} \cdot \hat{\bm{F}}^{e}$, where $\bm{M}^{VF}_{ij}$ is hydrodynamic mobility tensor relating the force acting on particle $j$ to the velocity of particle $i$. Similarly, $\hat{\bm{\Omega}}_i = \bm{M}^{\Omega F}_{ij} \cdot \hat{\bm{F}}^{e}$, where $\bm{M}^{\Omega F}_{ij}$ is the hydrodynamic mobility pseudotensor relating the force acting on particle $j$ to the angular velocity of particle $i$. Substituting these relations for the model flow quantities into \eqref{RT3} yields 
\begin{align} \label{RT4}
  \hat{\bm{F}}^e \cdot \bm{V}_j  = \hat{\bm{F}}^e \cdot \sum_{i} \left(\bm{F}^{e}_i \cdot \bm{M}^{VF}_{ij} + \bm{L}^e_i \cdot \bm{M}^{\Omega F}_{ij} +   \int_{S_i} \bm{v}_s \bm{n} : \K_j(\bm{x}) \,\rmd S\right).
\end{align}

Similarly, we introduce another set of $n$ model problems with flow fields  ($\hat{\bm{v}}^{L}_j(\bm{x})$, $\hat{\bm{\sigma}}^{L}_j(\bm{x})$), with $j \in \{1,2,3, \dots, n\}$, in which a external torque $\hat{\bm{L}}^e$ acts only on the $j$th particle, the remaining particles are torque-free, and all particles (including particle $j$) are force-free.  Analogously to \eqref{RT4}, we obtain 
\begin{align} \label{RT5}
\hat{\bm{L}}^e \cdot \bm{\Omega}_j  = \hat{\bm{L}}^e \cdot \sum_i \left( \bm{F}^{e}_i \cdot \bm{M}^{VL}_{ij} + \bm{L}^e_i \cdot \bm{M}^{\Omega L}_{ij} + \int_{S_i} \bm{v}_s \bm{n} : \N_j(\bm{x})\, \rmd S\,\right),
\end{align}
where $\bm{M}^{VL}_{ij}$ (respectively $\bm{M}^{\Omega L}_{ij}$) is the hydrodynamic mobility pseudotensor (tensor) relating a torque applied to particle $j$ to the velocity (angular velocity) of particle $i$ in the model problem. Also, $\N_{j}(\bm{x})$ is the corresponding rank-3 stress-propagating pseudotensor field, defined so that $\hat{\bm{\sigma}}^L_j(\bm{x}) = \N_j(\bm{x}) \cdot \hat{\bm{L}}^e$. 

Because \eqref{RT4} and \eqref{RT5} hold for arbitrary $\hat{\bm{F}}^e$ and $\hat{\bm{L}}^e$, respectively, we obtain the velocity and angular velocity of particle  $j$ as
\begin{subequations}\label{RTVel1}
\begin{align} 
\bm{V}_j  &= \sum_i \left( \bm{F}^{e}_i \cdot \bm{M}^{VF}_{ij} + \bm{L}^e_i \cdot \bm{M}^{\Omega F}_{ij} + \int_{S_i} \bm{v}_s \bm{n} : \K_j(\bm{x})\, \rmd S\,\right) \\
\bm{\Omega}_j  &= \sum_i \left( \bm{F}^{e}_i \cdot \bm{M}^{VL}_{ij} + \bm{L}^e_i \cdot \bm{M}^{\Omega L}_{ij} + \int_{S_i} \bm{v}_s \bm{n} : \N_j(\bm{x})\, \rmd S\,\right).
\end{align}
\end{subequations}
For the common case of active particles free of external forces and torques, \eqref{RTVel1} reduces to 
\begin{subequations}\label{RTVel2}
\begin{align}
\bm{V}_j  &= \sum_i  \int_{S_i} \bm{v}_s \bm{n} : \K_j(\bm{x})\, \rmd S\,, \\
\bm{\Omega}_j  &= \sum_i  \int_{S_i} \bm{v}_s \bm{n} : \N_j(\bm{x})\, \rmd S\,.
\end{align}
\end{subequations}

The expressions \eqref{RTVel1} and \eqref{RTVel2} are exact for any $n$-particle system under Stokes flow with arbitrary surface velocity distributions on the particles [\eqref{RTVel2} assumes force- and torque-free particles] and are central results of this work. The advantage of the current approach is that a direct computation of the flow generated due to the surface slip $\bm{v}_s(\bm{x} \in S_i)$ is not necessary to determine the motion of the particles. The relevant hydrodynamic quantities are the mobility tensors and the surface stress propagator fields $\bm{K}_j(\bm{x} \in S_i)$ and $\bm{N}_j(\bm{x} \in S_i)$, which depend only on the configurations of the particles and are independent of the surface slip velocity.  We recognize that these tensors are not trivial to compute, although analytical and numerical methods developed in recent years makes their approximation feasible. The mobility matrices, in particular, have been discussed extensively in the literature, for example, using Stokesian dynamics \citep{dur87_dynamic,bra88_stokesian}, multipole methods \citep{maz82_manysphere, ekiel09_multipole}, and simple point-particle approximations \citep{hocking1964behaviour,ekiel2006spherical,metzger2007falling}. We also note that the presence of external no-slip or free-surface boundaries is accounted for directly through the mobilities \citep{bha05_multipole_walls, swa07_particles, swa10_parallel} and the stress propagators.

\section{Widely separated spheres} \label{SecSpheres}

In the remainder of this paper, we consider particles free of external forces or torques, and focus on the ``active'' contribution due to the surface velocity distributions $\bm{v}_s(\bm{x} \in S_i)$. While numerical techniques such as boundary integral methods are necessary for more complex shapes, analytical methods can be used to approximate the hydrodynamic tensors  for spherical particles. Below, we consider widely separated spheres with radii $a_i$, and introduce for later convenience a characteristic particle radius $a$ and a characteristic separation distance $d = D a$, where $D \gg 1$ is a dimensionless separation distance. 

In order to use \eqref{RTVel2}, we first develop approximations for the stress propagators $\K_j(\bm{x})$ and $\N_j(\bm{x})$ defined by the model problems. Our strategy will be to use the method of reflections, where the ``incident'' flow due to particle $j$ is ``reflected'' by the other particles $i \neq j$ to satisfy boundary conditions on all the surfaces. For our purposes, it will suffice to consider only zeroth and first reflections though iterative applications of the method results in improved approximations to $\K_j(\bm{x})$ and $\N_j(\bm{x})$. Because the particles are force-free, velocities decay at least as fast as $r^{-2}$, so we expect that a small number of terms in such an expansion are effective at approximating the dynamics.

We introduce the notation $f^{(k)}$ to refer to the approximation to a quantity $f$ at the level of the $k$th reflection. The leading approximation to the fluid velocity, $\hat{\bm{v}}^{(0)}(\bm{x})$, is that produced by the action of either a force or a torque (depending on the model problem) acting on particle $j$ in isolation. Near a distant particle $i \neq j$, this leading-order flow can be expressed using a Taylor series (about $\bm{x} = \bm{x}_i$) as
\begin{equation} \label{Vinf}
  \hat{\bm{v}}^{(0)}(\bm{x}) = \hat{\bm{U}}^{(0)}_i + \frac{1}{2}\hat{\bm{\omega}}^{(0)}_i \wedge \bm{r}_i + \bm{r}_i \cdot \hat{\bm{E}}^{(0)}_i + \bm{r}_i \bm{r}_i \cdot \hat{\bm{G}}^{(0)}_i + \dots\,,
\end{equation}
where we recall that $\bm{r}_i = \bm{x} - \bm{x}_i$. Here, $\hat{\bm{U}}^{(0)}_i = \hat{\bm{v}}^{(0)}|_{\bm{x}_i}$, $\hat{\bm{\omega}}^{(0)}_i = \nabla \wedge \bm{v}^{(0)}|_{\bm{x}_i}$, $\hat{\bm{E}}^{(0)}_i = \frac{1}{2} \left\{\nabla \bm{v}^{(0)} + \left(\nabla \bm{v}^{(0)}\right)^T\right\}\big|_{\bm{x}_i}$ and $\hat{\bm{G}}^{(0)}_i = \frac{1}{2} \nabla \nabla \bm{v}^{(0)}\big|_{\bm{x}_i}$ as, respectively, the velocity, vorticity, rate-of-strain tensor and one half the curvature tensor of the zeroth reflection flow at $\bm{x} = \bm{x}_i$. 

Particle $i$, when exposed to the velocity field \eqref{Vinf}, produces a disturbance flow (the first reflection) that decays away from it. Here, we are only interested in the surface traction $\bm{n} \cdot \hat{\bm{\sigma}}|_{S_i}$, since the stress propagators $\K_j(\bm{x})$ and $\N_j(\bm{x})$ are evaluated on particle surfaces in \eqref{RTVel1} and \eqref{RTVel2}. Using known results for linear flows  \citep{kim_karirila_book,leal07_book}, the surface traction on $S_i$, associated with the first reflection, can be expressed as 
\begin{align} \label{Traction}
\bm{n} \cdot \hat{\bm{\sigma}}^{(1)}\big|_{S_i} = -\frac{3 \mu}{2 a_i} \left(\hat{\bm{V}}_i -  \hat{\bm{U}}_i^{(0)} \right)- 3 \mu \left(\hat{\bm{\Omega}}_i - \frac{1}{2}\hat{\bm{\omega}}_i^{(0)}\right) \wedge \bm{n} + 5 \mu\, \bm{n} \cdot \hat{\bm{E}}_i^{(0)} + \bm{n} \cdot \hat{\bm{\sigma}}^{(1)}_{\bm{G}}\big|_{S_i}, \quad i \neq j.
\end{align}
Here, $\hat{\bm{\sigma}}^{(1)}_{\bm{G}}\big|_{S_i}$ is surface traction contributed by the quadratic moment $\hat{\bm{G}}^{(0)}_i$, and can be directly evaluated using equation (3.9) of Nadim and Stone \cite{nad91_quadratic}, see the Appendix (note that $\hat{\bm{G}}^{(0)}_i$ here is identical to $\bm{K}$ in \citep{nad91_quadratic}). 

Since particle $i \neq j$ is force- and torque-free in the model problem, we obtain (either from direct integration of \eqref{Traction} or through Fax\'{e}n's laws) that $\hat{\bm{V}}_i -  \hat{\bm{U}}_i^{(0)} = (a_i^2/3) (\bm{I} : \hat{\bm{G}}^{(0)}_i)$ and $\hat{\bm{\Omega}}_i - \frac{1}{2}\hat{\bm{\omega}}_i^{(0)} = \bm{0}$. Consequently, we obtain 
\begin{equation} \label{FreeParticleTraction}
  \bm{n} \cdot \hat{\bm{\sigma}}^{(1)}\big|_{S_i} =  5 \mu\, \bm{n} \cdot \hat{\bm{E}}_i^{(0)} - \frac{\mu a_i}{2} \bm{I} : \hat{\bm{G}}^{(0)}_i + \bm{n} \cdot \hat{\bm{\sigma}}^{(1)}_{\bm{G}}\big|_{S_i}, \quad i \neq j,
\end{equation}
with corrections decaying as $D^{-4}$. Thus,  $\hat{\bm{E}}_i^{(0)}$ and $\hat{\bm{G}}_i^{(0)}$ in the model problems let us directly obtain leading approximations for $\K_j(\bm{x})$ and $\N_j(\bm{x})$ on the particle surfaces $i \neq j$, including hydrodynamic interactions up to the force-quadrupole and torque-quadrupole, respectively. The traction on particle $j$ (on which the external force or torque acts in the model problem) is directly obtained from the zeroth reflection flow, up to relative errors of $O(D^{-5})$ or smaller due to reflections from other spheres.

The tensors $\hat{\bm{E}}_i^{(0)}$ and $\hat{\bm{G}}_i^{(0)}$ in the model problems can be computed using gradients of the Stokeslet. From the set of model problems involving forces, we obtain (see the Appendix)
\begin{align}  \label{StressPropagatorK}
\bm{n} \cdot \bm{K}_j \big|_{S_i} = \left\{
\begin{array}{ll}
   \displaystyle -\frac{\bm{I}}{4 \pi a_i^2} + O(a^{-2} D^{-5}), & i = j \\[15pt]
   \displaystyle \frac{5}{8 \pi} \bm{n} \cdot \left(\frac{\bm{I}}{r_{ji}^3} - \frac{\bm{r}_{ji} \bm{r}_{ji}}{r_{ji}^5} \right) \bm{r}_{ji} + \frac{5 a_i}{16 \pi} \left\{\frac{3}{4} \frac{\bm{I}}{r_{ji}^3} - \frac{1}{2} \frac{\bm{n} \bm{n}}{r_{ji}^3} - \frac{23}{4} \frac{\bm{r}_{ji} \bm{r}_{ji}}{r_{ji}^5} - \frac{11}{2} \left(\bm{n} \cdot \bm{r}_{ji} \right) \frac{\bm{n} \bm{r}_{ji} + \bm{r}_{ji} \bm{n}}{r_{ji}^5} \right.  \\[12pt]
  \displaystyle \left.\qquad  + \frac{1}{4} \left(\bm{n} \cdot \bm{r}_{ji}\right)^2\left(\frac{\bm{I}}{r_{ji}^5} +105 \frac{\bm{r}_{ji} \bm{r}_{ji}}{r_{ji}^7} \right) - 2 \frac{\left(\bm{r}_{ji} \wedge \bm{n}\right)\left(\bm{r}_{ji} \wedge \bm{n}\right)}{r_{ji}^5} \right\} + O(a^{-2}D^{-4}), &  i \neq j,
\end{array}\right.
\end{align}
where $\bm{r}_{ji} = \bm{x}_i - \bm{x}_j$ and $r_{ji} = |\bm{r}_{ji}|$. From model problems involving a torque on particle $j$,  we find (see the Appendix)
\begin{align}  \label{StressPropagatorN}
\bm{n} \cdot \bm{N}_j \big|_{S_i} = \left\{
\begin{array}{ll}
   \displaystyle -\frac{3}{8 \pi a_i^3}\, \bm{\epsilon} \cdot \bm{n} + O(a^{-3} D^{-6}), & i = j \\[15pt]
   \displaystyle -\frac{15}{16 \pi} \frac{\bm{r}_{ji} \left(\bm{r}_{ji} \wedge \bm{n}\right) +  \bm{\epsilon}\cdot \bm{r}_{ji} (\bm{r}_{ji} \cdot \bm{n}) \,}{r_{ji}^5} + O(a^{-3} D^{-4}), &  i \neq j\,,
\end{array}\right.
\end{align}
where $\bm{\epsilon}$ is the permutation tensor. Observe that we have retained only linear velocity gradients to determine $\bm{N}_j$, although quadratic terms (which yield terms of $O(a^{-3} D^{-5})$ for $i \neq j$) can be included similarly to \eqref{StressPropagatorK}. 

Substituting the above relations into \eqref{RTVel2} yields the velocity and angular velocity of the particles in terms of a given surface velocity distribution as
\begin{subequations}\label{RT_Farfield}
\begin{align} 
\bm{V}_j  &\approx -\frac{1}{4 \pi a_j^2}\int_{S_j} \bm{v}_s \,\rmd S \;+\;  \frac{5}{8 \pi} \sum_{i, \,i \neq j}  \bm{r}_{ji} \left(\frac{\bm{I}}{r_{ji}^3} - 3 \frac{\bm{r}_{ji} \bm{r}_{ji}}{r_{ji}^5} \right) : \left(\int_{S_i} \bm{v}_s \bm{n}\, \rmd S \right)  \nonumber \\ 
 &\qquad + \frac{5 }{64 \pi} \sum_{i, i \neq j} a_i \int_{S_i} \bm{v}_s \cdot \left\{3\frac{\bm{I}}{r_{ji}^3} - 2 \frac{\bm{n} \bm{n}}{r_{ji}^3}- 23 \frac{\bm{r}_{ji} \bm{r}_{ji}}{r_{ji}^5} - 22 \left(\bm{n} \cdot \bm{r}_{ji} \right) \frac{\bm{n} \bm{r}_{ji} + \bm{r}_{ji} \bm{n}}{r_{ji}^5} \right. \nonumber \\
 &\qquad\qquad \qquad \qquad \qquad \left.+  \left(\bm{n} \cdot \bm{r}_{ji}\right)^2\left(\frac{\bm{I}}{r_{ji}^5} +105 \frac{\bm{r}_{ji} \bm{r}_{ji}}{r_{ji}^7} \right) - 8 \frac{\left(\bm{r}_{ji} \wedge \bm{n}\right)\left(\bm{r}_{ji} \wedge \bm{n}\right)}{r_{ji}^5} \right\} \rmd S \label{RT_Farfield_V}\\[15pt]
\bm{\Omega}_j &\approx -\frac{3}{8 \pi a_j^3}\int_{S_j} \bm{n} \wedge \bm{v}_s\, \rmd S  -\frac{15}{16 \pi} \sum_{i, i \neq j} \frac{1}{r_{ji}^5} \int_{S_i} \left( \left(\bm{n} \cdot \bm{r}_{ji}\right)\left(\bm{r}_{ji} \wedge \bm{v}_s\right)  + \left(\bm{v}_s \cdot \bm{r}_{ji}\right)\left(\bm{r}_{ji} \wedge \bm{n}\right)\right) \rmd S \label{RT_Farfield_Omega}, 
\end{align}
\end{subequations}
up to terms of $O(V_s (n-1) D^{-4})$ for velocity and $O(V_s a^{-1} (n-1) D^{-4})$ for angular velocity, $V_s$ being the characteristic scale of $\bm{v}_s$ and $n$ being the total number of particles. The first integral of \eqref{RT_Farfield_V} is the self-propulsion speed of a single sphere with surface slip, derived by Stone and Samuel \cite{sto96_PRL_squirmer}, the second corresponds to stresslet interactions between spheres, while the third integral is due to interactions of force quadrupoles (including source dipoles). Observe that the leading interaction terms fall off as $D^{-2}$, reflecting the fact that the leading hydrodynamic interactions between the (force- and torque-free) particles are due to stresslet flows. Similarly, the first integral of \eqref{RT_Farfield_Omega} represents the rotation of an isolated sphere with a surface velocity $\bm{v}_s(\bm{x})$ \citep{sto96_PRL_squirmer}, while the second integral includes interactions due to torque-dipoles (force-quadrupoles) and decays as $D^{-3}$. The effects of external forces and torques can be described by linear superposition of \eqref{RT_Farfield} with results for passive no-slip particles [see \eqref{RTVel1}], which may be obtained, for instance, by Stokesian dynamics \citep{bra88_stokesian} or multipole methods \citep{ekiel09_multipole}.

\section{Examples of swimming with hydrodynamic interactions}
The expressions \eqref{RT_Farfield} describe a system of $n$ widely separated force- and torque-free spheres with arbitrary surface slip velocity distributions. Below, we discuss some examples to demonstrate the applicability of the theory to a few common situations.

\subsection{Swimming with a prescribed multipolar surface velocity}
We first consider, as a relatively general case, the interactions of swimmers with prescribed surface velocity distributions. In many situations, the surface velocity is itself expressed in terms of a multipole expansion, e.g.,
\begin{equation} \label{Vs_multipole}
  \bm{v}_s(\bm{x}) = \left(\bm{I} - \alpha_i \bm{n} \bm{n}\right) \cdot \bm{\lambda}_i + \left(\bm{I} - \beta_i \bm{n} \bm{n}\right) \cdot \bm{B}_i \cdot \bm{n}    \qmbox{for} \bm{x} \in S_i.
\end{equation}
Here, $\bm{\lambda}_i$ (a vector) and $\bm{B}_i$ (a rank-2 tensor) are, respectively, dipolar and quadrupolar strengths of the surface velocity distribution (both having units of velocity), and $\alpha_i$ and $\beta_i$ are dimensionless scalar parameters. An isolated swimmer swims with a velocity in proportion to $\bm{\lambda}_i$, but is independent of $\bm{B}_i$, while its rotation rate is linear in $\bm{B}_i$ and independent of $\bm{\lambda}_i$; see  \citep{sto96_PRL_squirmer}. Hydrodynamic interactions modify this feature:  using \eqref{RT_Farfield}, and noting that $\int_{S_i} n_{\alpha} n_{\beta}  \rmd S = \frac{4}{3}\pi a_i^2 \delta_{\alpha \beta}$ and $\int_{S_i} n_{\alpha} n_{\beta} n_{\gamma} n_{\delta}\rmd S =  \frac{4}{15} \pi a_i^2 \left(\delta_{\alpha \beta} \delta_{\gamma \delta} + \delta_{\alpha \gamma} \delta_{\beta \delta} + \delta_{\alpha \delta} \delta_{\gamma \beta}\right)$ [Greek subscripts denote Cartesian indices], we find 
\begin{subequations}\label{V_Omega_Multi}
\begin{align} 
\bm{V}_j &= -\left(1- \frac{\alpha_j}{3}\right)\bm{\lambda}_i + \frac{5}{6} \sum_{i, \,i \neq j}  a_i^2 \left(1 - \frac{2\beta_i}{5}\right)\bm{r}_{ji} \left(\frac{\bm{I}}{r_{ji}^3} - 3 \frac{\bm{r}_{ji} \bm{r}_{ji}}{r_{ji}^5} \right) : \bm{B}_i + \frac{1}{3} \sum_{i, \,i \neq j}  \alpha_i a_i^3 \bm{\lambda}_i \cdot \left(\frac{\bm{I}}{r_{ji}^3} - 3 \frac{\bm{r}_{ji} \bm{r}_{ji}}{r_{ji}^5} \right), \\
\bm{\Omega}_j &= -\frac{1}{2 a_j} \bm{\epsilon}:\bm{B}_j - \frac{5}{4} \sum_{i, i \neq j} \frac{a_i^2}{r_{ji}^5} \left(1 - \frac{2 \beta_i}{5}\right)\; \bm{r}_{ji} \wedge \left\{ \left(\bm{B}_i + \bm{B}_i^T\right) \cdot \bm{r}_{ji}\right\}\,.
\end{align}
\end{subequations}
We observe that both the translation and rotation of the particles depend on the $\bm{B}_i$ of other particles due to hydrodynamic interactions. The last term of (\ref{V_Omega_Multi}a) couples the velocities of all the particles to the surface motions of the other particles. We also note that the next multipole (an octupole, characterized by a rank-3 tensor) will modify the translation velocity both through the isolated-swimmer contribution and force-quadrupolar hydrodynamic interactions. In many systems involving orientable particles such as swimming phytoplankton, rod-like bacteria or synthetic Janus colloids, multipole moments of the surface velocity are tied to particle orientation, introducing an additional degree of coupling between rotation and translation. 

\subsection{Autophoresis of n spheres with surface sorption}

Here, we show how the relations \eqref{V_Omega_Multi} directly apply to the phoretic motion of chemically active colloids. A suspension of catalytic colloids may generate concentration gradients of a chemical species in the fluid as a result of surface reactions. Combined with sufficient geometric asymmetry, the colloids may move autonomously via diffusiophoresis \citep{michelin13,twosphere,wall,var18_clustering_autophoresis}. In such systems, provided that the interactions with the solute occur on short length scales (typically a molecular length scale or for charged systems, the Debye layer thickness), the surface slip velocity takes a form that involves the chemical concentration $c$, 
\begin{equation} \label{Vs_DP}
    \bm{v}_s(\bm{x}) = -\Gamma(c) \,  (\bm{I} - \bm{n} \bm{n}) \cdot \nabla c \qmbox{for} \bm{x} \in S_i\,,
\end{equation}
where $\Gamma(c)$ is a $c$-dependent mobility. Note that in an externally applied concentration gradient $\nabla c^{\infty}$, isolated particles translate with velocity $\Gamma(c) \nabla c^{\infty}$ according to \eqref{V_Omega_Multi}, i.e. particles with positive mobility move towards higher concentrations. 

For non-electrolytic solutes, $\Gamma$ is often independent of $c$, whereas for polar solvents and charged particles, the mobility is typically of the form $\Gamma(c) = \Lambda c^{-1}$, where $\Lambda$ is a diffusiophoretic mobility that depends on the charge of the particle and the properties of the solvent \citep{derjaguin47,anderson82,anderson,JFM1984}.  We consider the limit of small P\'{e}clet number, $V d/D_s \ll 1$, where $V$ is the characteristic particle velocity and $D_s$ is the molecular diffusivity of the solute, and assume a quasi-static adjustment of the concentration field to the motion of the particles (valid on time scales $t \gg d^2/D_s$). Then, for particles that produce solute at their surfaces with a specified flux $j(\bm{x})$, the solute concentration is governed by 
\begin{subequations}
\begin{align}
  \nabla^2 c &= 0 \qmbox{in the fluid volume, with} \\
   - D_s \bm{n} \cdot \nabla c &= j(\bm{x}) \qmbox{on the particle surfaces.}
\end{align}
\end{subequations}

The concentration field can be constructed by the method of reflections for widely separated spheres (see \citep{var18_clustering_autophoresis} for a more detailed discussion). We denote the solution to the transport problem around particle $j$ in isolation by $c^{(0)}_j(\bm{x})$. Then, particle $i$ appears immersed in a concentration field produced by a superposition of the individual $c^{(0)}_j(\bm{x})$, $j \neq i$, and produces a disturbance field (the first reflection) to maintain the flux condition on its surface. Retaining terms up to quadrupolar order, the concentration field around particle $i$ can be expressed as 
\begin{subequations} \label{Conc}
\begin{align}
    c(\bm{x}) &\approx c^{\infty}+ c^{(0)}_i(\bm{x}) + \sum_{j \neq i} c_j^{(0)}(\bm{x}_i) + \bm{r}_i \cdot \bm{d}^{(0)}_i \left(1 + \frac{1}{2}\frac{a_i^3}{r_i^3} \right) +  \bm{r}_i \bm{r}_i : \bm{Q}^{(0)}_i \left(1 + \frac{2}{3}\frac{a_i^5}{r_i^5} \right) + \dots , \qmbox{where} \\
    \bm{d}^{(0)}_i &= \sum_{j \neq i} \nabla c_j^{(0)}(\bm{x}_i), \qmbox{and} \bm{Q}^{(0)}_i = \sum_{j \neq i} \frac{1}{2}\nabla \nabla c_j^{(0)}(\bm{x}_i), 
\end{align}
\end{subequations}
and $c^{\infty}$ is the uniform ambient concentration far away from all of the particles. Substituting  \eqref{Conc} into \eqref{Vs_DP} lets us evaluate the slip velocity distribution on $S_i$. 

For a uniform surface flux over the particle surface [$j(\bm{x} \in S_i) = j_i$)], we have  $c_{i}^{(0)} = c^{\infty} J_i a_i/r_i$, where $J_i = j_i a_i/(D_s c^{\infty})$ is the dimensionless flux on the surface of $i$. In this case, evaluating  \eqref{Conc} and substituting it into \eqref{Vs_DP} yields
\begin{equation} \label{Vslip_autophoresis}
  \bm{v}_s(\bm{x}) \approx - \Gamma\left(c (\bm{x})\right) \;(\bm{I} - \bm{n} \bm{n}) \cdot  \left\{\frac{3}{2} \bm{d}^{(0)}_i + \frac{10}{3}  \bm{Q}^{(0)}_i \cdot \bm{n}\right\} \qmbox{for} \bm{x} \in S_i.
\end{equation}
Approximating $c({\bm{x} \in S_i}) \approx c^{\infty}+ c^{(0)}_i(\bm{x} \in S_i) + \sum_{j,j\neq i} c^{(0)}_j(\bm{x}_i)$ in the evaluation of $\Gamma$, we observe that \eqref{Vslip_autophoresis} is of the form discussed in \eqref{Vs_multipole}. Thus, we immediately obtain
\begin{align} \label{V_autophoresis}
  \bm{V}_i &= \Gamma \left(c|\bm{x} \in {S_i}\right) \left\{\bm{d}^{(0)}_i - \frac{5}{3}   \sum_{j, j \neq i}  a_j^2 \bm{r}_{ij} \left(\frac{\bm{I}}{r_{ij}^3} - 3 \frac{\bm{r}_{ij} \bm{r}_{ij}}{r_{ij}^5} \right) \cdot \bm{Q}^{(0)}_i - \frac{1}{2} \sum_{j, \,j \neq i}  a_j^3 \bm{d}_j^{(0)} \cdot \left(\frac{\bm{I}}{r_{ij}^3} - 3 \frac{\bm{r}_{ij} \bm{r}_{ij}}{r_{ij}^5} \right)\right\},
\end{align} 
where 
\begin{equation}
  \bm{d}_i^{(0)} = - c^{\infty} \sum_{k, k \neq i} \frac{J_k a_k \bm{r}_{ki}}{r_{ki}^3} \qmbox{and} \bm{Q}_i^{(0)} = -\frac{c^{\infty}}{2} \sum_{k, k \neq i} J_k a_k^2 \left(\frac{\bm{I}}{r_{ki}^3} - 3 \frac{\bm{r}_{ki} \bm{r}_{ki}}{r_{ki}^5} \right).
\end{equation}
The above expressions recover the results of \citep{var18_clustering_autophoresis}, but do so without requiring multiple reflections between the spheres. We note that the leading interactions, decaying as $D^{-2}$, are due to concentration dipoles, whereas hydrodynamic interactions decay as $D^{-5}$.

We do not analyze the hydrodynamic interactions further, but present some example results below, exploring the leading-order influence of chemical inhomogeneities on the trajectories of particles. For electrolyte diffusiophoresis, $\Gamma(c) = \Lambda_i/c$, which results in the sphere velocities
\begin{align} \label{V_diffusio}
    \bm{V}_i =   \frac{\Lambda_i}{1 + J_i}\sum_{j,\, j\neq i} J_j a_j \frac{\bm{r}_{ij}}{r_{ij}^3}. 
\end{align}
Equation \eqref{V_diffusio} is the generalization of the two-particle result obtained in previous work \citep{yan19_autophoresis}. The factor of $\Lambda_i/(1 + J_i)$ is replaced by a constant $\Gamma$ in the case of non-electrolyte diffusiophoresis, for which $\Gamma$ is typically independent of $c$. Note that the interactions in \eqref{V_diffusio} are not hydrodynamic but rather occur through the concentration field. 

\subsubsection{Three spheres}
Drawing from the work of Hocking \citep{hocking1964behaviour} for the sedimentation of clusters of spheres, we analyze some aspects of the autophoresis of three particles in the absence of hydrodynamic interactions. The velocity of each particle, with possibly distinct values of radii $a_i$, dimensionless diffusiophoretic mobility $\Lambda_i$ and dimensionless chemical flux $J_i$, according to (\ref{V_diffusio}), is
\begin{subequations}
\begin{align}
\frac{\rmd \bm{x}_1}{\rmd t}=\frac{\Lambda_1}{1+J_1}\left(J_2a_2\frac{\bm{x}_2-\bm{x}_1}{|\bm{x}_2-\bm{x}_1|^3}+J_3a_3\frac{\bm{x}_3-\bm{x}_1}{|\bm{x}_3-\bm{x}_1|^3}\right),   \\
\frac{\rmd \bm{x}_2}{\rmd t}=\frac{\Lambda_2}{1+J_2}\left(J_1a_1\frac{\bm{x}_1-\bm{x}_2}{|\bm{x}_1-\bm{x}_2|^3}+J_3a_3\frac{\bm{x}_3-\bm{x}_2}{|\bm{x}_3-\bm{x}_2|^3}\right),   \\
\frac{\rmd \bm{x}_3}{\rmd t}=\frac{\Lambda_3}{1+J_3}\left(J_1a_1\frac{\bm{x}_1-\bm{x}_3}{|\bm{x}_1-\bm{x}_3|^3}+J_2a_2\frac{\bm{x}_2-\bm{x}_3}{|\bm{x}_2-\bm{x}_3|^3}\right).   
\end{align}
\end{subequations}
The centroid of the triangle formed by the particles, which has a position $\bm{x}_c = \frac{1}{3} \sum_i \bm{x}_i$, then evolves as 
\begin{equation}
    \dd{\bm{x}_c}{t} = \left(\frac{\Lambda_1 J_2 a_2}{1 + J_1} - \frac{\Lambda_2 J_1 a_1}{1 + J_2}\right) \frac{\bm{r}_{12}}{3 r_{12}^3}  + \left(\frac{\Lambda_2 J_3 a_3}{1 + J_2} - \frac{\Lambda_3 J_2 a_2}{1 + J_3}\right) \frac{\bm{r}_{23}}{3 r_{23}^3} + \left(\frac{\Lambda_3 J_1 a_1}{1 + J_3} - \frac{\Lambda_1 J_3 a_3}{1 + J_1}\right) \frac{\bm{r}_{31}}{3 r_{31}^3}.
\end{equation}
The centroid remains stationary over time if the properties of all three particles are related by 
\begin{equation}
 \frac{\Lambda_i}{a_i J_i (1+J_i)} = \mbox{constant}, \quad i \in \{1,2,3\}.
\end{equation}
As a special case of the above expression, the centroid of the triangle formed by identical particles is stationary.

The vector surface area of the triangle $\bm{\Delta}$ can be defined by $2\bm{\Delta}=\bm{r}_{12}\wedge\bm{r}_{23}$. Then, the rate of change of $\bm{\Delta}$ is
\begin{eqnarray}
2\frac{\rmd \bm{\Delta}}{\rmd t}&=&\left[\frac{\rmd}{\rmd t}(\bm{x}_2-\bm{x}_1)\right]\wedge(\bm{x}_3-\bm{x}_2)+(\bm{x}_2-\bm{x}_1)\wedge\frac{\rmd}{\rmd t}(\bm{x}_3-\bm{x}_2)\nonumber\\
&=&\left[\left(\frac{\Lambda_2 J_1a_1}{1+J_2}+\frac{\Lambda_1 J_2a_2}{1+J_1}\right)\frac{\bm{r}_{21}}{r_{21}^3}-\frac{\Lambda_1J_3a_3}{1+J_1}\frac{\bm{r}_{13}}{r_{13}^3}\right]\wedge\bm{r}_{23} + \bm{r}_{12}\wedge\left[\left(\frac{\Lambda_2 J_3a_3}{1+J_2}+\frac{\Lambda_3 J_2a_2}{1+J_3}\right)\frac{\bm{r}_{32}}{r_{32}^3}+\frac{\Lambda_3 J_1a_1}{1+J_3}\frac{\bm{r}_{31}}{r_{31}^3}\right]. \label{delt}
\end{eqnarray}
Since $2\bm{\Delta}=\bm{r}_{12}\wedge\bm{r}_{23} = \bm{r}_{23} \wedge \bm{r}_{31} = \bm{r}_{31} \wedge \bm{r}_{12}$, 
equation (\ref{delt}) can be simplified as
\begin{align}
\frac{\rmd \bm{\Delta}}{\rmd t}=-\Bigg[\frac{1}{r_{12}^3}\left(\frac{\Lambda_1J_2a_2}{1+J_1} + \frac{\Lambda_2 J_1a_1}{1+J_2}\right)+\frac{1}{r_{23}^3}\left(\frac{\Lambda_2J_3a_3}{1+J_2}+\frac{\Lambda_3J_2a_2}{1+J_3}\right) +\frac{1}{r_{31}^3}\left(\frac{\Lambda_3J_1a_1}{1+J_3} + \frac{\Lambda_1J_3a_3}{1+J_1}\right)\Bigg]\bm{\Delta}.
\end{align}
If the three particles are identical, the area of triangle will either increase (when $\Lambda J < 0$) or decrease (when $\Lambda J > 0$) until the particles come into contact, noting that $1+J$ must be positive since the concentration field cannot be negative. This result can be understood by considering two identical particles, which only attract or repel each other. Note that the area is constant over time only in the trivial case where all particles have either zero mobility or zero flux.
\begin{figure}[t!]
  \includegraphics[scale=1]{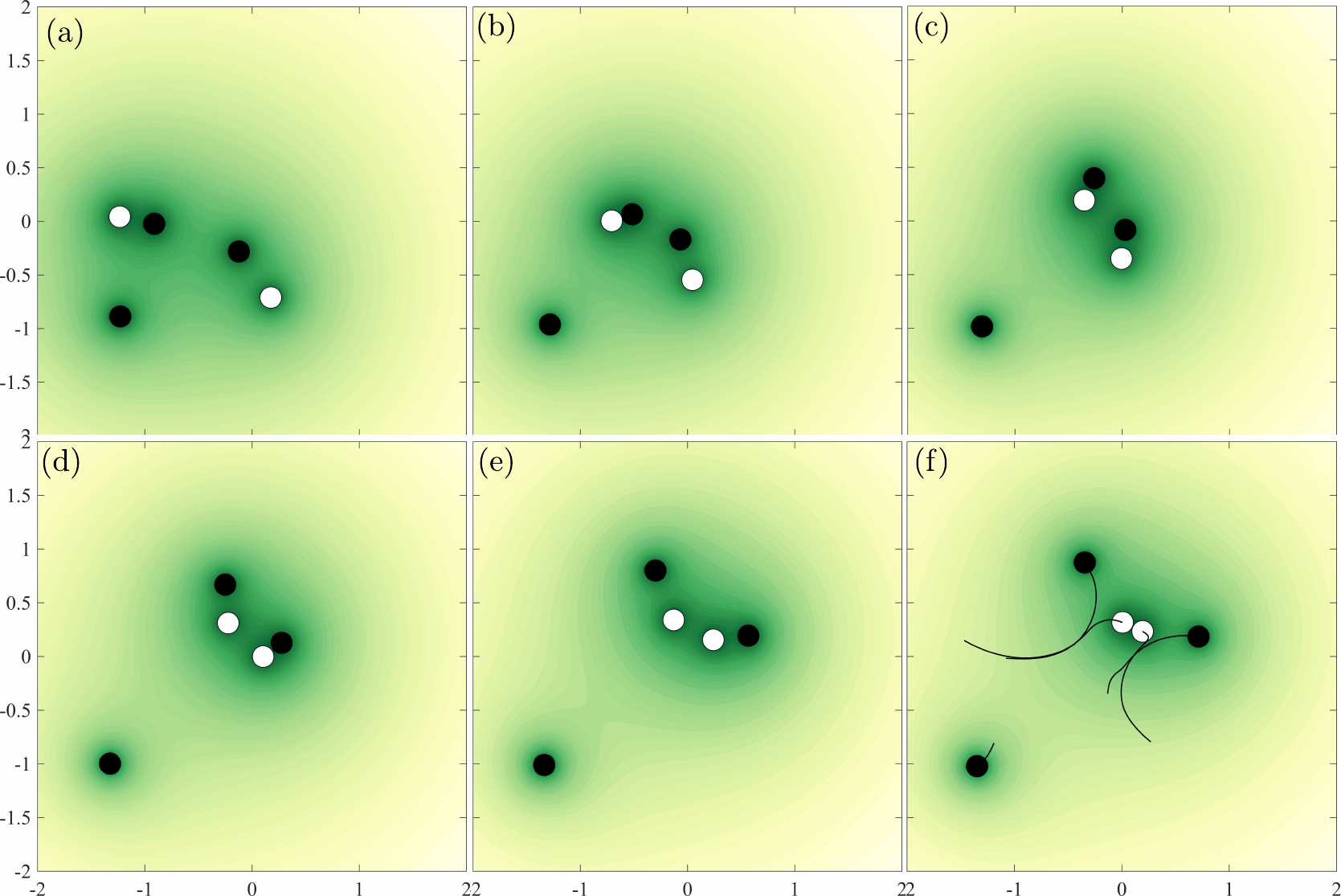}
\caption{Snapshots in time showing chasing and pair-switching of five spheres with uniform flux $J = 0.5$ and identical radius. Particles with positive $\Lambda$ are shaded white, and those with negative $\Lambda$ are shaded black; note that $|\Lambda|$ are equal for all the particles. Initially, two chaser pairs are established with negative heads and positive tails [panel (a)]. As the chasers run into each other (b,c), their heads repel and their tails attract (d,e). Consequently, the chaser pairs disintegrate and instead the two positive particles pair up, while the negative particles continually move outward towards regions of low concentration (f). Curves in panel (f) trace the paths of particle centers over time. }
\label{Fig5particles}
\end{figure}

\subsubsection{Numerical calculation of the autophoresis of 5 spheres}
Below, we consider spheres of identical radius $a$ and flux $J$, but with different mobility coefficients $\Lambda_i$.  We solve for the particle trajectories by numerically integrating \eqref{V_diffusio} in time, and apply a hard-core repulsive potential to prevent overlap of the particles. For $J>0$, the ion concentration increases around the particles, causing ``positive'' particles ($\Lambda_i > 0$), which move up concentration gradients, to attract each other, and ``negative'' particles ($\Lambda_i < 0$) to repel each other. The direction of the motion is reversed for $J < 0$, where the solute is depleted near the particles.

Upon contact of two particles with different mobilities, a particle $i$ will chase another particle $j$ if $\Lambda_i > \Lambda_j$. This behavior occurs due to a differential response of the particles to the same concentration gradient. Their combined velocity depends on their difference in mobility \citep{twosphere,yan19_autophoresis}, while the hard-core repulsion negates any attractive motion (proportional to their mean mobility). The end result is that for $J > 0$, the head of the chaser is negative relative to its tail.  


A simulation of five spheres, at different times, is shown in Fig. 2.  We find that particles often tend to pair off and form ``chasers'' based on their proximity and difference in mobility. However, these pairs are not permanent and can be disrupted as the chaser encounters other particles in its path. In Fig. \ref{Fig5particles}, two of the particles have positive mobility (white shading) and the other three have negative mobility of the same magnitude (black shading). Initially, two pairs of chasers are established. As time progresses, the chasers first approach each other and then turn away as their heads (negative particles) repel. This puts their tails (positive particles) in close proximity of each other, which subsequently attract. The result is that the chasers disintegrate, the positive particles form a relatively stationary pair, and the remaining particles move outward. 


\section{Conclusions} \label{SecConclusions}
We have developed, as our main result, a general formalism to describe the motion of hydrodynamically interacting active particles with arbitrary surface velocities. In contrast with earlier approaches, we use the Lorentz reciprocal theorem, which obviates the need to construct a detailed velocity field in the bulk fluid and instead relies on surface tractions in a model Stokes flow problem involving no-slip particles. We build on results in the literature to evaluate these surface tractions for the case of widely separated spheres, retaining contributions up to and including force and torque quadrupoles. We then apply the theory to example problems involving either purely hydrodynamic or a combination of chemical and hydrodynamic interactions. In the latter case we recover results of earlier work in the literature and demonstrate new effects due to chemical inhomogeneities in the system. 

Our results from section \ref{SecLRT} provide a general starting point to evaluate hydrodynamic interactions in active systems and equation \eqref{RT_Farfield} develops a versatile application of the theory to spherical particles. We emphasize the general nature of the method in that (i) it does not restrict the nature or the level of complexity of the surface velocity distribution, and (ii) it accounts for all hydrodynamic interactions up to the level of force quadrupoles. The general theory of section \ref{SecLRT} also remains applicable in the presence of boundaries.  These traits makes the present framework a versatile tool to describe hydrodynamic interactions in wide variety of active systems.


\appendix*
\section{Stress propagators in the model problems} \label{App:stress}
Here, we use known results from the literature to construct the stress propagators $\bm{K}_j$ and $\bm{N}_j$ introduced in section \ref{SecLRT}, accurate to the level of force-quadrupoles.  First, we  consider the set of model problems corresponding to a force acting on particle $j$, with no torque acting on any of the particles. The Stokeslet $\Jpf_j(\bm{x})$ and the corresponding stress propagator $\Kpf_j(\bm{x})$ for an external force acting at $\bm{x}_j$ in an unbounded medium are 
\begin{align} \label{Stokeslet}
 \Jpf_j(\bm{x}) = \frac{1}{8 \pi \mu} \left(\frac{\bm{I}}{r_j} + \frac{\bm{r}_j \bm{r}_j}{r_j^3} \right), \quad \Kpf_j(\bm{x}) = -\frac{3}{4 \pi} \frac{\bm{r}_j \bm{r}_j \bm{r}_j}{r_j^5},
\end{align}
where we recall that $\bm{r}_j = \bm{x} - \bm{x}_j$ (cf. Fig. \ref{FigSketch}).  Then, the leading approximation (zeroth reflection) to the velocity propagator in the model problem is $\bm{J}^{(0)}_j(\bm{x}) = \Jpf_j(\bm{x}) + \frac{a^2}{6} \nabla^2 \Jpf_j(\bm{x})$. The corresponding stress propagator is $\bm{K}^{(0)}_j(\bm{x}) = \Kpf_j(\bm{x}) + \frac{a^2}{6} \nabla^2 \Kpf_j(\bm{x})$. On the surface of particle $j$, the zeroth reflection gives the surface traction as $\bm{n} \cdot \bm{K}_j|_{S_j} \approx \bm{n} \cdot \bm{K}^{(0)}_j|_{S_j} = -\bm{I}/\left(4 \pi a_j^2\right)$, up to terms of $O(a^{-2} D^{-5})$.

A force- and torque-free particle $i \neq j$ is exposed to the ambient flow $\bm{v}^{(0)} = \bm{J}^{(0)}_j \cdot \bm{r}_j$, which is locally (around particle $i$) of the form \eqref{Vinf}. Neglecting the contribution due to the source dipole at $\bm{x}_j$, the quantities $\hat{\bm{E}}^{(0)}_i$ and $\hat{\bm{G}}^{(0)}_i$ are obtained by taking gradients of $\Jpf_j(\bm{x})$. For convenience we introduce the shorthand notation $\bm{R} = \bm{r}_{ji}$.  Using Greek indices to denote Cartesian components and employing the Einstein summation convention, we find
\begin{subequations} \label{EGForce}
\begin{align}
  \hat{E}^{F(0)}_{\alpha \beta} &= \frac{1}{8 \pi \mu} \left(\frac{\delta_{\alpha \beta}}{R} - \frac{R_{\alpha}R_{\beta}}{R^3}\right) R_{\delta}  \hat{F}^e_{\delta} \\
  \hat{G}^{F(0)}_{\alpha \beta \gamma} &= \frac{1}{16 \pi \mu} \left(-\frac{\delta_{\alpha \beta} \delta_{\gamma \delta}}{R^3} + 3 \frac{R_{\alpha} R_{\beta} \delta_{\gamma \delta}}{R^5} +\frac{\delta_{\beta \gamma} \delta_{\alpha \delta}}{R^3} - 3 \frac{r_{\beta} R_{\gamma} \delta_{\alpha \delta}}{R^5} + \frac{\delta_{\alpha \gamma} \delta_{\beta \delta}}{R^3} - 3 \frac{R_{\alpha} R_{\gamma} \delta_{\beta \delta}}{R^5} \right. \\
&\qquad - \left. \frac{3 R_{\delta}}{R^5} \left(\delta_{\alpha \beta} R_{\gamma} + \delta_{\beta \gamma} R_{\alpha} + \delta_{\alpha \gamma} R_{\beta} \right) + 15 \frac{R_{\alpha} R_{\beta} R_{\gamma} R_{\delta}}{R^7} \right) \hat{F}^e_{\delta},
\end{align}
\end{subequations}
where the superscript $F$ denotes that the above quantities are associated with the model problem involving a force on particle $j$. Using \eqref{FreeParticleTraction}, \eqref{EGForce} and the results for surface traction for a sphere in a general quadratic flow (equation (3.9) of \cite{nad91_quadratic}), we obtain the surface traction on the force- and torque-free particle $i \neq j$ as 
\begin{align} 
  \bm{n} \cdot \hat{\bm{\sigma}}_j^{F(1)} \big|_{S_i} &= \bigg[\frac{5}{8 \pi} \bm{n} \cdot \left(\frac{\bm{I}}{R^3} - \frac{\bm{R} \bm{R}}{R^5} \right) \bm{R} + \frac{5 a_i}{16 \pi} \left\{\frac{3}{4} \frac{\bm{I}}{R^3} - \frac{1}{2} \frac{\bm{n} \bm{n}}{R^3} - \frac{23}{4} \frac{\bm{R} \bm{R}}{R^5} - \frac{11}{2} \left(\bm{n} \cdot \bm{R} \right) \frac{\bm{n} \bm{R} + \bm{R} \bm{n}}{R^5} \right. \nonumber \\
  &\qquad \left. + \frac{1}{4} \left(\bm{n} \cdot \bm{R}\right)^2\left(\frac{\bm{I}}{R^5} +105 \frac{\bm{R} \bm{R}}{R^7} \right) - 2 \frac{\left(\bm{R} \wedge \bm{n}\right)\left(\bm{R} \wedge \bm{n}\right)}{R^5} \right\} \bigg]\cdot \hat{\bm{F}}^e.
\end{align}
This result directly leads to \eqref{StressPropagatorK}. 

We use a similar analysis for the set of model problems involving  a torque acting on sphere $j$, with all spheres being force-free. The leading approximation to the velocity due to the torque acting on sphere $j$ is  $\hat{\bm{v}}^{L (0)}(\bm{x}) = \left(\hat{\bm{L}}^{e} \wedge \bm{r}_j\right)/(8 \pi \mu r_j^3)$, where the superscript $L$ indicates that the model problem in question is one in which a torque acts on particle $j$.
Up to relative errors of $O(D^{-6})$, the traction on the surface of particle $j$ is determined by the zeroth reflection flow as $\bm{n} \cdot \hat{\bm{\sigma}}_j^{L}\big|_{\bm{x} \in S_j} \approx \bm{n} \cdot \hat{\bm{\sigma}}_j^{L(0)}\big|_{\bm{x} \in S_j} = 3/(8 \pi a_j^3) (\bm{n} \wedge \hat{\bm{L}}^{e})$. Taylor expanding $\hat{\bm{v}}^{L (0)}(\bm{x})$ about the center $\bm{x}_i$ of a particle $i \neq j$ identifies the rate-of-strain tensor as
\begin{align} \label{EL0}
   \hat{\bm{E}}^{L(0)}_i &= -\frac{3}{16 \pi \mu\, r_{ji}^5} \left(\bm{r}_{ji} \left(\hat{\bm{L}}^e \wedge \bm{r}_{ji}\right) + \left(\hat{\bm{L}}^e \wedge \bm{r}_{ji}\right)\bm{r}_{ji} \right).
\end{align}
Within the formalism of \eqref{FreeParticleTraction} one may include the contribution of $\hat{\bm{G}}^{L(0)}$ to the traction on $S_i$. However, this contribution results from a torque quadrupole (or part of a force octupole) and decays as $D^{-4}$, and so will be neglected here. Then, the surface traction on particle $i$, using \eqref{FreeParticleTraction}, is $\approx 5 \mu \bm{n} \cdot \hat{\bm{E}}_i^{L(0)}$, which results in \eqref{StressPropagatorN}.

\begin{acknowledgments}
The authors thank the National Science Foundation for support through the Center for Chemo-Mechanical Assembly (Grant No. NSF-CCI-1740630), and Lauren Zarzar for a stimulating seminar that helped motivate this work. BR thanks the Bourns College of Engineering, University of California, Riverside, for support through an initial complement. 
\end{acknowledgments}



\end{document}